\documentclass{appolb_xxx}
\usepackage{epsfig}

\newcommand{\eq}[1]{eq.~(\ref{#1})}

\newcommand{\Eq}[1]{Eq.~(\ref{#1})}
\newcommand{\Eqs}[2]{Eqs.(\ref{#1},\ref{#2})}

\newcommand{\ur}[1]{(\ref{#1})}

\newcommand{\beq}{\begin{equation}}
\newcommand{\eeq}{\end{equation}}

\newcommand{\la}[1]{\label{#1}}
\newcommand{\bea}{\begin{eqnarray}}
\newcommand{\eea}{\end{eqnarray}}
\newcommand{\ba}{\begin{array}}
\newcommand{\ea}{\end{array}}

\newcommand{\n}{\nonumber}
\newcommand{\Det}{{\rm Det}}

\begin{document}
\title{Statistical physics of dyons and confinement
\thanks{Presented at Cracow School of Theoretical Physics, June 13-22, 2008, Zakopane, Poland}
}
\author{Dmitri DIAKONOV
\address{Petersburg Nuclear Physics Institute, Gatchina 188300, St. Petersburg, Russia}
}
\maketitle

\begin{abstract}
We present a semiclassical description of the $SU(N)$ Yang--Mills theory
whose partition function at nonzero temperatures is approximated
by that of an ensemble of $N$ kinds of interacting dyons.
The ensemble is mathematically described by an exactly solvable
$3d$ quantum field theory, allowing calculation of correlations functions
relevant to confinement. We show that all known criteria of confinement
are satisfied in this semiclassical approximation: (i)~the average Polyakov
line is zero below some critical temperature, and nonzero above it,
(ii)~static quarks in any nonzero $N$-ality representation have linear
rising potential energy, (iii)~the average spatial Wilson loop falls off
exponentially with the area, (iv)~$N^2$ gluons are canceled out from the
spectrum, (v) the critical temperature is in good agreement with lattice data.
\end{abstract}
\PACS{11.15.-q,11.10.Wx,11.15.Tk}

\newpage

\tableofcontents

\newpage

\section{Philosophy}

Quantum Chromodynamics (QCD) is hardly an exactly solvable quantum field
theory, even in the large $N$ limit. Therefore, one can either do exact
calculations in a theory that has more symmetries but is not our world,
or work with QCD but make approximations. The first is useful as a
theoretical laboratory, the second is necessary to understand semi-quantitatively
the key phenomena, to explain experimental data, and to make predictions.

An approximation is considered to be legitimate if there is a systematic way
of improving its accuracy. The semiclassical approach belongs to this category.
One chooses a saddle-point classical field and then has to take into account
quantum fluctuations about it. Part of the fluctuations are ultra-violet
and are thus the same as in empty space. Therefore their role is to renormalize
the bare coupling constant; at this point the famous dimensional transmutation
occurs, when the ultraviolet cutoff in a proper combination with the bare
coupling constant forms the QCD scale parameter $\Lambda$, the only dimensional
scale that henceforth will be in the theory. What is left, is a series in the
't Hooft {\it running} coupling $\lambda\equiv N\alpha_s/2\pi$ coming from
loop expansion in the background of classical configurations.

The argument of the running coupling $\lambda$ is determined by the largest
scale in the vacuum, $max(T,\,n^{\frac{1}{4}})$, where $T$ is temperature,
and $n$ is the mean $4d$ density of the (large) classical field configurations.
For example, near the deconfinement temperature $T\approx T_c\approx \Lambda$
the running coupling is approximately
$\lambda=\frac{11}{3}\ln\left(4\pi T/e^{\gamma_{\rm E}}\right)\approx\frac{1}{7}$~\cite{DO}.
The numerically large factor $2\pi$ in the argument of the logarithm is not
accidental but related to the fact that it is actually not the temperature itself
but rather the Matsubara frequency $2\pi T$ that defines the scale. At zero temperature
many QCD specialists believe that $\alpha_s$ does not grow above the value
of 0.5, which gives $\lambda\approx\frac{1}{4}$. Therefore, in the whole range
of temperatures within the confining phase the semiclassical approximation
is expected to yield the accuracy of 15-25\%, already in the 1-loop approximation
(provided the saddle point is chosen correctly!) with a possibility for rapid
improvement when higher loops are taken into account. We shall see, however,
that the actual accuracy can be much better than this estimate. It is not
a too big price to pay for solving the most challenging riddle in 35 years: confinement.

We shall be considering the pure Yang--Mills theory based on the $SU(N)$ gauge
group in a broad range of temperatures between 0 and $T_c$, the deconfinement
phase transition temperature. Although the formalism we use is designed for
nonzero $T$, we shall see that the physical observables we find (such as the string
tension) have a finite limit when $T\to 0$. In this limit the nonzero temperature
can be thought of as an infrared regulator. After all, our world's temperature is
$2.7\,{\rm K}\neq 0$.

Confinement, as we understand it today and learn from lattice experiments with
a pure glue theory, has in fact many facets, and all have to be explained. Let us
enumerate the main:
\begin{itemize}
\item the average Polyakov line in any $N$-ality nonzero representation
of the $SU(N)$ group is zero below $T_c$ and nonzero above it
\item the potential energy of two static colour sources (defined through
the correlation function of two Polyakov lines) asymptotically rises linearly
with the separation; the slope called the string tension depends only on
the $N$-ality of the sources
\item the average of the spatial Wilson loop decays exponentially with
the area spanning the contour; at vanishing temperatures the spatial (``magnetic'') string
tension has to coincide with the ``electric'' one, for all representations
\item the mass gap: no massless gluons left in the spectrum.
\end{itemize}

These properties have been obtained in Ref.~\cite{DP-07} with Victor Petrov, which
is the base for this presentation.

\section{Yang--Mills theory at nonzero temperatures}

The Yang--Mills (YM) partition function can be written as a path or functional
integral over the spatial components of the connection $A_i(t,{\bf x})$ satisfying
the periodic boundary conditions up to a gauge transformation $\Omega({\bf x})$
over which one has to integrate separately~\cite{GPY}:
\bea\n
{\cal Z}&=&\sum_{\rm gauge\;invariant\; states}
\langle n\left| e^{-\beta {\cal H}}\right|n\rangle
=\int \, D\Omega({\bf x})DA_i({\bf x})\qquad\qquad \left[\beta=\frac{1}{T}\right]\\
\la{Z1}
&\times &
\int_{A_i({\bf x})}^{A_i({\bf x})^{\Omega({\bf x})}}\!DA_i(t,{\bf x})
\,\exp\left[\!-\frac{1}{g^2}\,\int_0^\beta\!dt\,\int\!d^3{\bf x}\,\Tr\left(\dot A_i\dot A_i
\!+\!B_iB_i\right)\right]
\eea
where $B_i=\epsilon_{ijk}\left(\partial_jA_k-\frac{i}{2}[A_jA_k]\right)$ is the
magnetic field strength and $A_i^\Omega\equiv\Omega^\dagger A_i \Omega+i\Omega^\dagger\partial_i\Omega$
is the gauge-transformed potential. $A_i,\,B_i$ are $N\!\times\!N$ matrices belonging to the
$su(N)$ algebra while $\Omega$ is an element of the $SU(N)$ group.

One can rewrite the partition function in a more customary form by introducing gauge-transformed
integration variables $A_i$ that are strictly periodic in time, and trading $\Omega$ for
the time component of the YM potential $A_4$ that is also periodic:
\bea\la{Z2}
{\cal Z}&=&\int\!DA_\mu(t,{\bf x})\,\exp\left(-\frac{1}{2g^2}\!\int_0^{\frac{1}{T}}\!dt
\int\!d^3{\bf x}\,\Tr F_{\mu\nu}F_{\mu\nu}\right),\\
\n
&&A_\mu\left(t+\frac{1}{T},{\bf x}\right)=A_\mu(t,{\bf x}),
\eea
where $F_{\mu\nu}$ is the usual field strength. This form stresses the fact that
Euclidean $O(4)$ symmetry is restored as $T\to 0$.

An important variable is the Polyakov loop: in the formulation \ur{Z2} it is
the path-ordered exponent
\beq
L({\bf x})={\cal P}\,\exp\left(i\int_0^{\frac{1}{T}}\!dt\,A_4(t,{\bf x})\right)
\qquad \left(=\Omega({\bf x})\right).
\la{L}\eeq
In the formulation \ur{Z1} it is nothing but the $SU(N)$ matrix $\Omega({\bf x})$
over which there is a final integration in \Eq{Z1}. The eigenvalues of $L({\bf x})$
are gauge invariant; we parameterize them as
\beq
L={\rm diag}\left(e^{2\pi i \mu_1},e^{2\pi i \mu_2},\ldots ,e^{2\pi i \mu_N}\right),
\qquad \mu_1\!+\!\ldots\!+\mu_N=0,
\la{mu}\eeq
and assume that the phases of these eigenvalues are ordered:
$\mu_1\leq\mu_2\leq\ldots\leq\mu_N\leq\mu_{N+1}\equiv\mu_1\!+\!1$.
We shall call the set of $N$ phases $\{\mu_m\}$ the ``holonomy'' for short.
Apparently, shifting $\mu$'s by integers does not change the eigenvalues, hence all
quantities have to be periodic in all $\mu$'s.

The holonomy is said to be ``trivial'' if $L$ belongs to one of the $N$ elements
of the group center $Z_N$. For example, in $SU(3)$ the three trivial holonomies
are
\bea\n
1. && \mu_1=\mu_2=\mu_3=0\quad\Longrightarrow\quad
L=\left(\begin{array}{ccc}1&0&0\\
0&1&0\\ 0 & 0 & 1\end{array}\right)\,,\\
\n
2. && \mu_1=-\frac{2}{3},\,\mu_2=\frac{1}{3},\,\mu_3=\frac{1}{3}\quad\Longrightarrow\quad
L=e^{\frac{2\pi i}{3}}\left(\begin{array}{ccc}1&0&0\\
0&1&0\\ 0 & 0 & 1\end{array}\right)\,,\\
\n
3. && \mu_1=-\frac{1}{3},\,\mu_2=-\frac{1}{3},\,\mu_3=\frac{2}{3}\quad\Longrightarrow\quad
L=e^{-\frac{2\pi i}{3}}\left(\begin{array}{ccc}1&0&0\\
0&1&0\\ 0 & 0 & 1\end{array}\right)\,.
\eea
Trivial holonomy corresponds to equal $\mu$'s, {\it modulo} unity. Out of all possible
combinations of $\mu$'s a distinguished role is played by equidistant $\mu$'s:
\beq
\mu_m^{\rm conf}=-\frac{1}{2}-\frac{1}{2N}+\frac{m}{N},\qquad \Tr L=0.
\la{muconf}\eeq
For example, in $SU(3)$ it is
\beq
\mu_1=-\frac{1}{3},\,\mu_2=0,\,\mu_3=\frac{1}{3}\Longrightarrow
L=\left(\begin{array}{ccc}e^{-\frac{2\pi i}{3}}&0&0\\
0&e^{\frac{0\,\pi i}{3}}&0\\ 0 & 0 & e^{\frac{2\pi i}{3}}\end{array}\right),\qquad \Tr L=0.
\la{muconf3}\eeq
We shall call it ``maximally non-trivial'' or ``confining'' holonomy as it
corresponds to $\Tr L=0$ which the 1$^{\rm st}$ confinement requirement.

Immediately, an interesting question arises: Imagine we take the YM partition function,
be it in form \ur{Z1} or \ur{Z2}, and integrate out all degrees of freedom except
the eigenvalues $\{\mu_m\}$ of the Polyakov loop $L({\bf x})$ (or $\Omega({\bf x})$),
which, in addition, we take slowly varying in space. What is the effective action
for $\mu$'s? What set of $\mu$'s is preferred dynamically by the YM system of fields?

In general, it is a difficult calculational problem that can be addressed using various
approximations but in one case the result is known exactly. It is the case of
the supersymmetric ${\cal N}\!=\!1$ version of the YM theory (SYM) where in addition to gluons
there are gluinos in the adjoint representation. In order not to spoil supersymmetry one
takes not the real temperature but rather a $4d$ space compactified in the time
direction, $R^3\!\times\!S^1$. The difference is that in the ``real temperature''
case one uses periodic conditions in the Euclidean time direction for boson fields
(gluons) and {\it anti}periodic conditions for fermion fields (gluinos) -- that spoils
supersymmetry; in the ``compactification'' case one implies periodic conditions for
both kinds of fields, which supports supersymmetry. However, we shall anyway call
the inverse circumference of the compactified time direction ``temperature'' for short.

There is no perturbative contribution to the potential energy in question as function of
$\mu$'s (directly related in this case to the holomorphic superpotential) because of the
supersymmetric cancelation between boson and fermion loops, and the only contribution is
nonperturbative coming from dyons. It can be reliably computed in the limit of high
``temperatures'' and then claimed to be actually independent of temperature owing to
the holomorphy typical in supersymmetry. The result~\cite{DHKM-99} is that the potential
energy of the system has the minimum at precisely the ``maximally non-trivial'' or
``confining'' holonomy \ur{muconf}.

In the non-supersymmetric pure YM theory, there is a perturbative effective action for
slowly varying $\mu$'s. It can be understood as gluon loop(s) in the background
of a slowly varying field $A_4({\bf x})$. The effective action can be expanded in
the number of gradients of $\mu$'s. The zero-order term, the potential energy with no
derivatives, has been computed long ago in Refs.~\cite{GPY,NW}:
\beq
P^{\rm pert}=V\,\left.\frac{(2\pi)^2T^3}{3}\sum_{m>n}^N
(\mu_m-\mu_n)^2[1-(\mu_m-\mu_n)]^2\right|_{{\rm mod}\;1},
\la{Ppert}\eeq
Since the piece with no derivatives implies constant $\mu$'s, it has to be proportional
to the 3-volume $V$, and hence to $T^3$ by dimensions. $P^{\rm pert}$ has exactly $N$
zero minima when all $\mu$'s are equal {\it modulo} unity. Hence, $P^{\rm pert}$ says
that at high temperatures the system prefers one of the $N$ trivial holonomies
corresponding to the Polyakov loop being one of the $N$ elements of the center $Z_N$,
see Fig.~1, top.

\begin{figure}[h]
\includegraphics[width=0.48\textwidth]{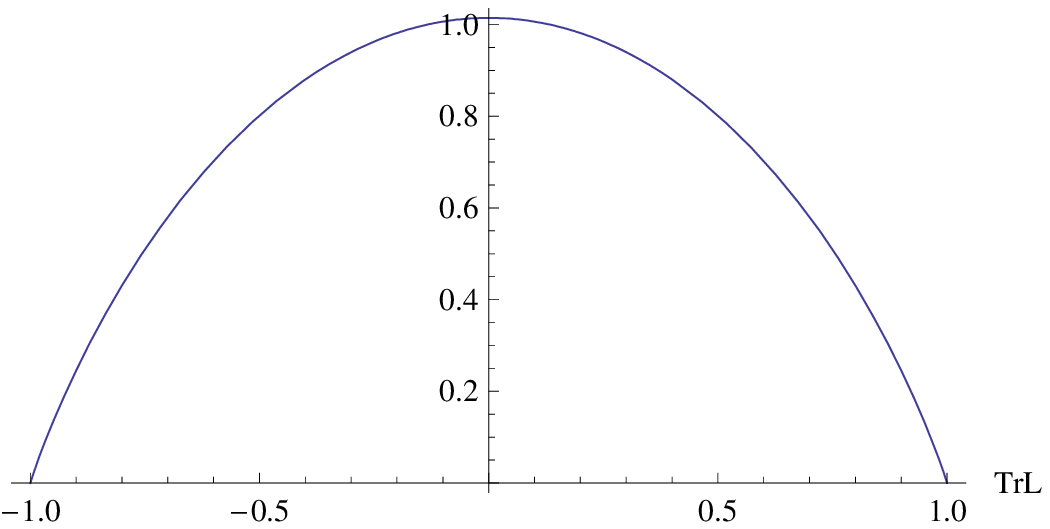} 
\includegraphics[width=0.48\textwidth]{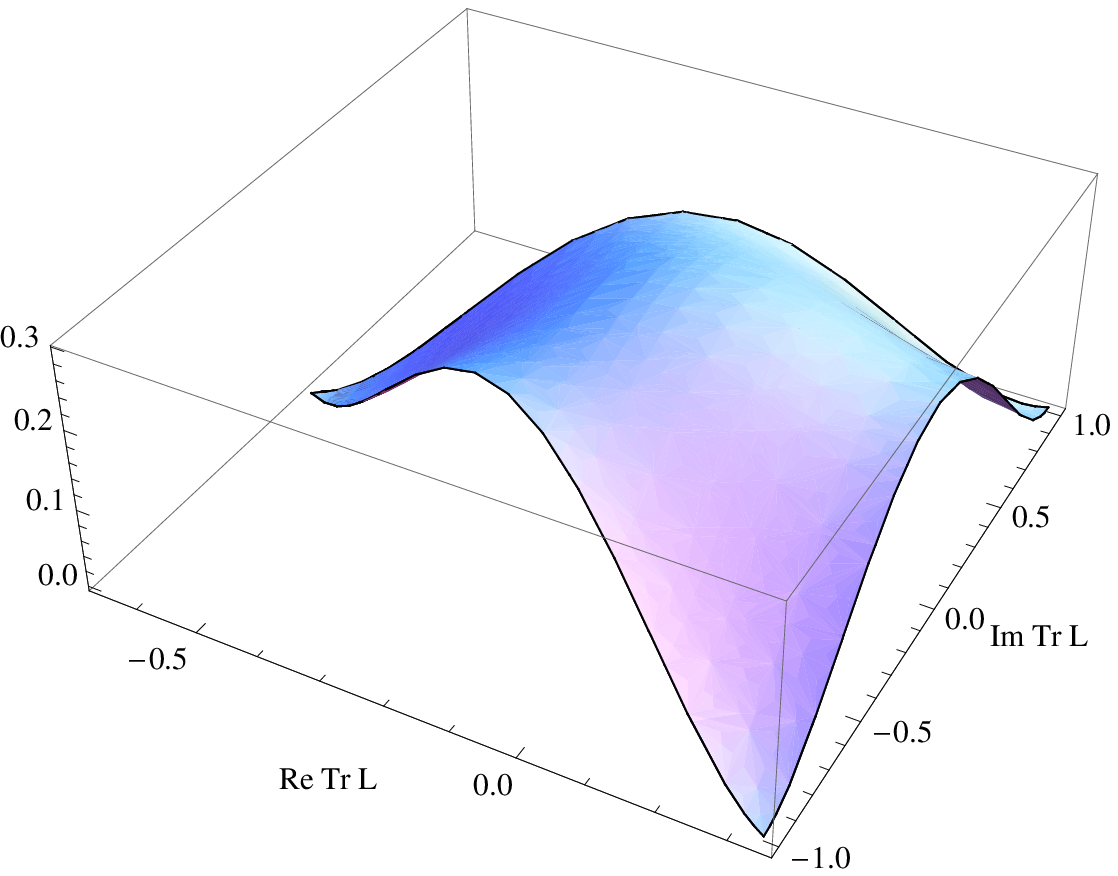} 
\includegraphics[width=0.48\textwidth]{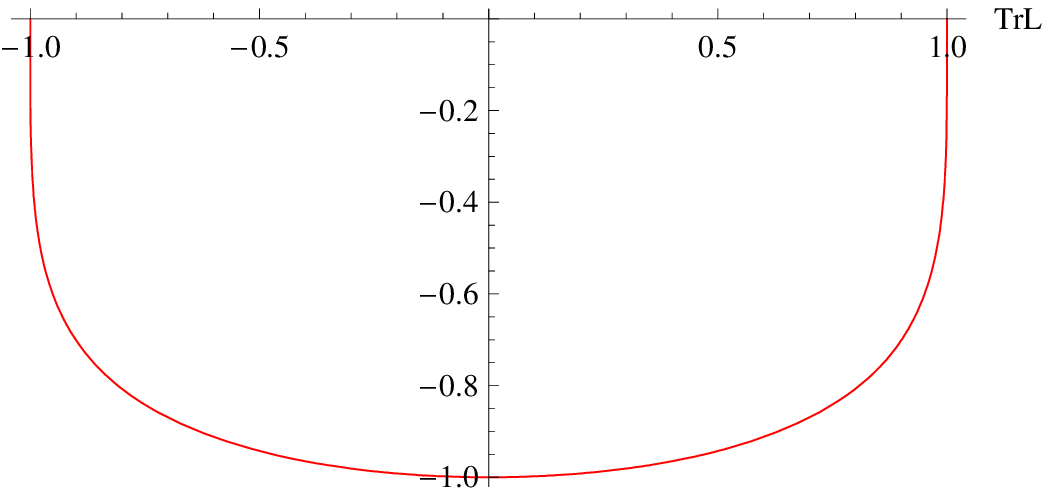}  
\includegraphics[width=0.48\textwidth]{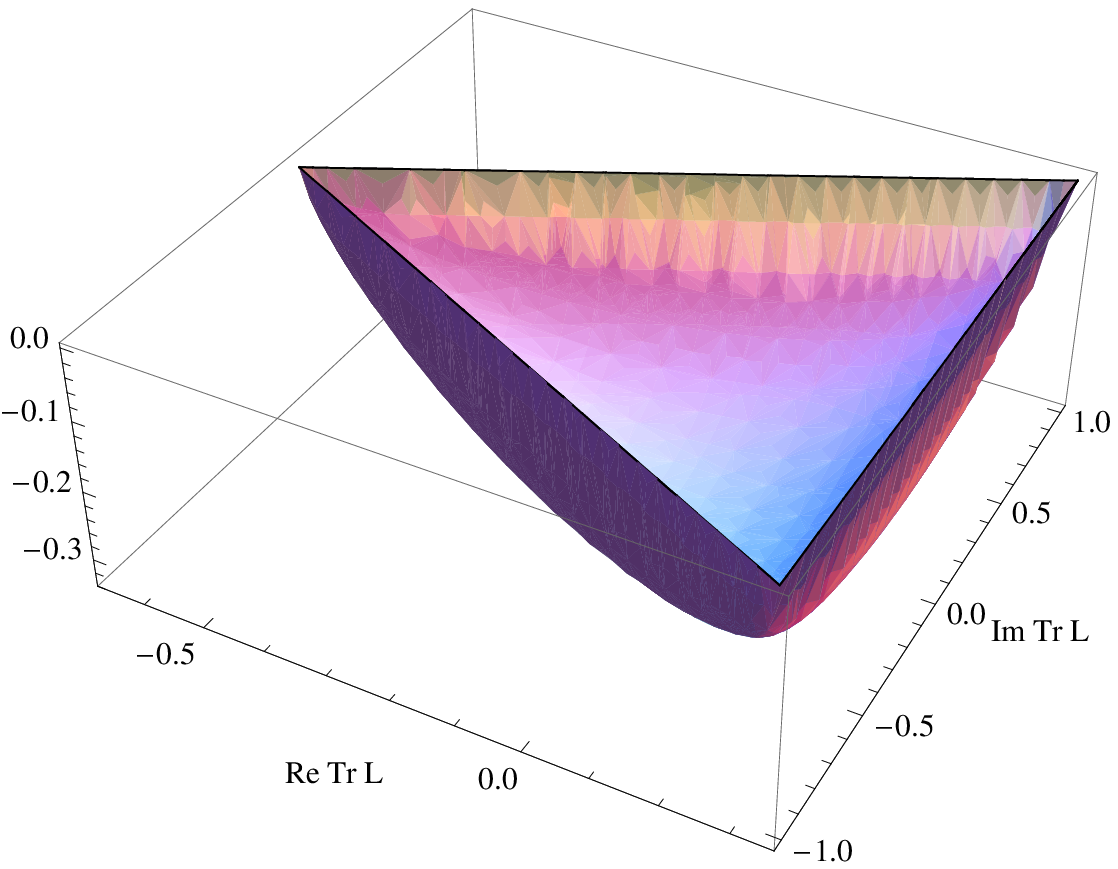}  
\caption{The perturbative ({\it top}) {\it vs.} nonperturbative ({\it bottom}) potential energy
as function of the Polyakov line for the $SU(2)$ ({\it left}) and $SU(3)$ ({\it right}) groups.
The perturbative potential energy has minima where the Polyakov loop is one of the $N$ elements
of the center $Z_N$ and is maximal at the ``confining'' holonomy. The nonperturbative
potential energy, on the contrary, has a single and non-degenerate minimum at the confining
holonomy corresponding to $\Tr L=0$.}
\end{figure}

However, gradient terms in the effective action indicate that there is
a problem with the trivial-holonomy points, already at the perturbative level.
Indeed, the two-derivative term is~\cite{DO}
\bea\n
S_{\rm 2-der}^{\rm pert}&=&\sum_{m>n}^N
\int\!d^3x\left[\partial_i(\mu_m\!-\!\mu_n)\right]^2\frac{11}{12}T
\left[H(\mu_m\!-\!\mu_n)+2\,\log\left(\frac{4\pi T}{\Lambda\,e^{\gamma_E}}\right)\right],\\
&&H(\nu)=\left[\psi(\nu)+\psi(1-\nu)+2\gamma_E\right]_{\rm mod\; 1},\quad
\psi(\nu)=\frac{d}{d\nu}\ln \Gamma(\nu).
\la{2der}\eea
Since $\psi(\nu)\approx -1/\nu$ at small $\nu$, the gradient term becomes
negative near ``trivial'' holonomy, which signals its instability even in perturbation
theory.

We shall show below that a semiclassical configuration -- an ensemble of dyons with quantum
fluctuations about it -- generates a nonperturbative free energy shown in Fig.~1, bottom.
It has the opposite behaviour of the perturbative potential energy, having
the minimum at the equidistant (confining) values of the $\mu$'s. There is a fight between
the perturbative and nonperturbative contributions to the free energy~\cite{D-02}.
Since the perturbative contribution is $\!\sim\!T^4$ with respect to the nonperturbative one,
it certainly wins when temperatures are high enough, and the system is then forced into one
of the $N$ vacua thus breaking spontaneously the $Z_N$ symmetry. At low temperatures the
nonperturbative contribution prevails forcing the system into the confining vacuum.
This is the mechanism of the confinement-deconfinement phase transition. It is of the second
order for $N\!=\!2$ but first order for $N\!=\!3$ and higher, in agreement with lattice
findings.

\section{Dyon saddle points}

Dyons or Bogomolny--Prasad--Sommerfield (BPS) monopoles~\cite{BPS} are (anti) self-dual solutions
of the nonlinear Maxwell equations, $D_\mu^{ab}F^b_{\mu\nu}=0$. In $SU(N)$ there are exactly $N$
kinds of `fundamental' dyons with Coulomb asymptotics for both electric and magnetic fields
(hence the term ``dyon''):
\beq
\pm{\bf E}={\bf B}\stackrel{|{\bf x}|\!\to\!\infty}{=}\frac{1}{2}\frac{{\bf x}}{|{\bf x}|^3}
\times\left(\begin{array}{ccc}
1&0&0\\0&-1&0\\0&0&0\end{array}\right),
\left(\begin{array}{ccc}
0&0&0\\0&1&0\\0&0&-1\end{array}\right),
\left(\begin{array}{ccc}
-1&0&0\\0&0&0\\0&0&1\end{array}\right).
\la{EB}\eeq
Dyon solutions are labeled by the holonomy or the set of $\mu$'s at spatial infinity:
\beq
A_4(|{\bf x}|\!\to\!\infty)\to 2\pi T
\left(\begin{array}{ccc}
\mu_1&0&0\\0&\mu_2&0\\0&0&\mu_3\end{array}\right)
\la{A4}\eeq
(we illustrate it for the case of $SU(3)$). The explicit expressions for
the solutions in various gauges can be found {\it e.g.} in the Appendix of Ref.~\cite{DP-SUSY}.
Inside the cores which are of the size $\sim 1/(T(\mu_{m+1}-\mu_m))$, the fields are large,
nonlinearity is essential. The action density is time-independent everywhere and is
proportional to the temperature. Isolated dyons are thus $3d$ objects but with finite
action $S_{\rm dyon}=(8\pi^2/g^2)(\mu_{m+1}-\mu_m)$ independent of temperature
(here $\mu_{N+1}\equiv\mu_1+1$).
The full action of all $N$ kinds of well-separated dyons together is that of one standard
instanton: $S_{\rm inst}=8\pi^2/g^2$.

In the semiclassical approach, one has first of all to find the statistical weight with which
a given classical configuration enters the partition function. It is given by $\exp(-{\rm Action})$,
times the determinant$^{-1/2}$ from small quantum oscillations about the saddle point. For
an isolated dyon as a saddle-point configuration, this factor diverges linearly in the infrared
region owing to the slow Coulomb decrease of the dyon field \ur{EB}. It means that isolated
dyons are not acceptable as saddle points: they have zero weight, despite finite classical
action. However, one may look for classical solutions that are superpositions of $N$ fundamental
dyons, with zero net magnetic charge. The small-oscillation determinant must be infrared-finite
for such classical solutions, if they exist.

\section{Instantons with non-trivial holonomy}

Remarkably, the needed classical solution has been found a decade ago by Kraan
and van Baal~\cite{KvB} and independently and simultaneously by Lee and Lu~\cite{LL},
see also~\cite{LeeYi}. We shall call them for short the ``KvBLL instantons''; an
alternative name is ``calorons with nontrivial holonomy''. The solution was first found
for the $SU(2)$ group but soon generalized to the arbitrary $SU(N)$~\cite{KvBSUN}.
A nice overview of the solutions has been presented by Pierre van Baal at the 2003
School in Zakopane~\cite{vBZak}. We shall mention only the essentials here.

The general solution depends on Euclidean time $t$ and space ${\bf x}$ and is parameterized
by $3N$ positions of $N$ kinds of `constituent' dyons in space ${\bf x}_1,\ldots, {\bf x}_N$
and their $U(1)$ phases $\psi_1,\ldots,\psi_N$. All in all, there are $4N$ collective
coordinates characterizing the solution (called the moduli space), of which the action
$S_{\rm inst}=8\pi^2/g^2$ is in fact independent, as it should be for a general
solution with a unity topological charge. The solution also depends explicitly on
temperature $T$ and on the holonomy $\mu_1,\ldots,\mu_N$:
\beq
A_\mu^{\rm KvBLL}=\bar A_\mu^a(t, {\bf x};\, {\bf x}_1,\ldots, {\bf x}_N,\psi_1,\ldots,\psi_N;\,
T,\mu_1,\ldots,\mu_N).
\la{KvBLLi}\eeq
The solution is a relatively simple expression given by elementary functions.
If the holonomy is trivial (all $\mu$'s are equal {\it modulo} unity) the expression
takes the form of the strictly periodic $O(3)$ symmetric caloron~\cite{HS} reducing further
to the standard $O(4)$ symmetric BPST instanton~\cite{BPST} in the $T\to 0$ limit.
At small temperatures but arbitrary holonomy, the KvBLL instanton also has only a small
${\cal O}(T)$ difference with the standard instanton.

One can plot the action density of the KvBLL instanton in various corners of the
parameter (moduli) space, see Fig.~2.

\begin{figure}[htb]
\includegraphics[width=0.32\textwidth]{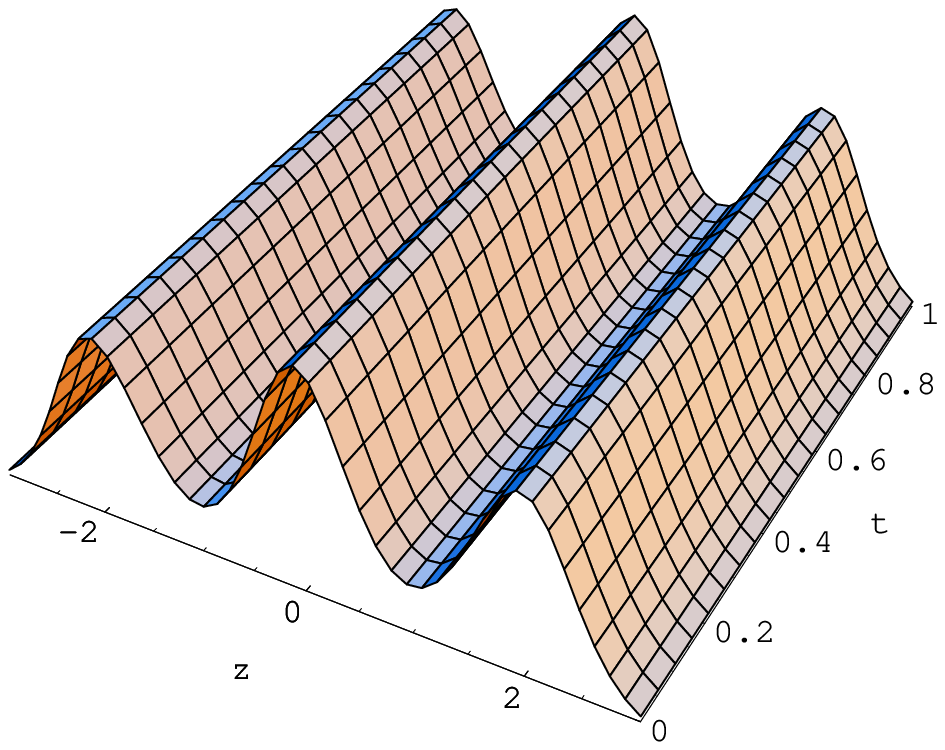} 
\includegraphics[width=0.32\textwidth]{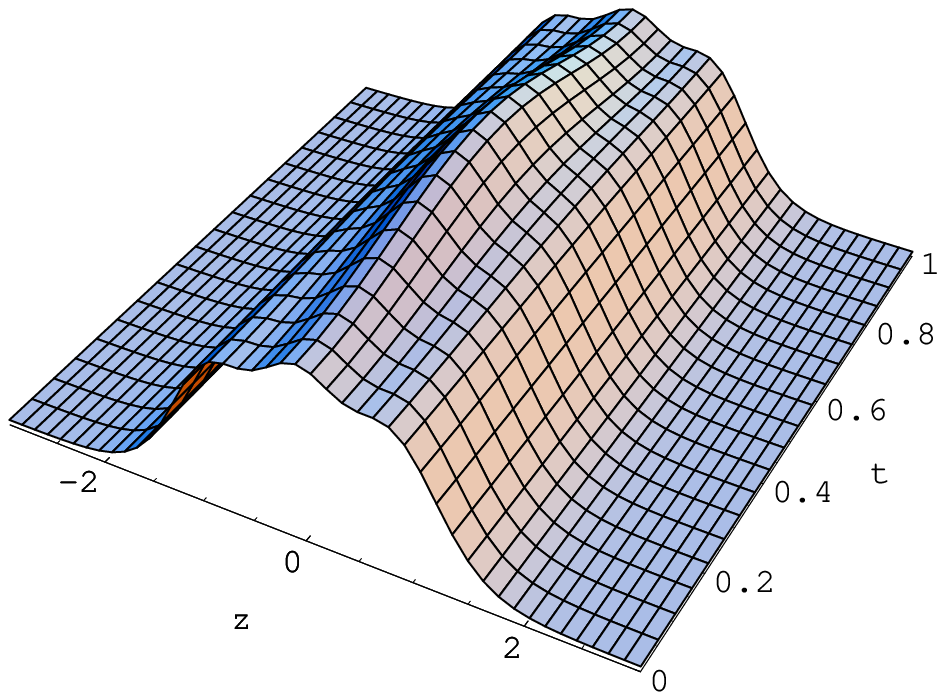} 
\includegraphics[width=0.32\textwidth]{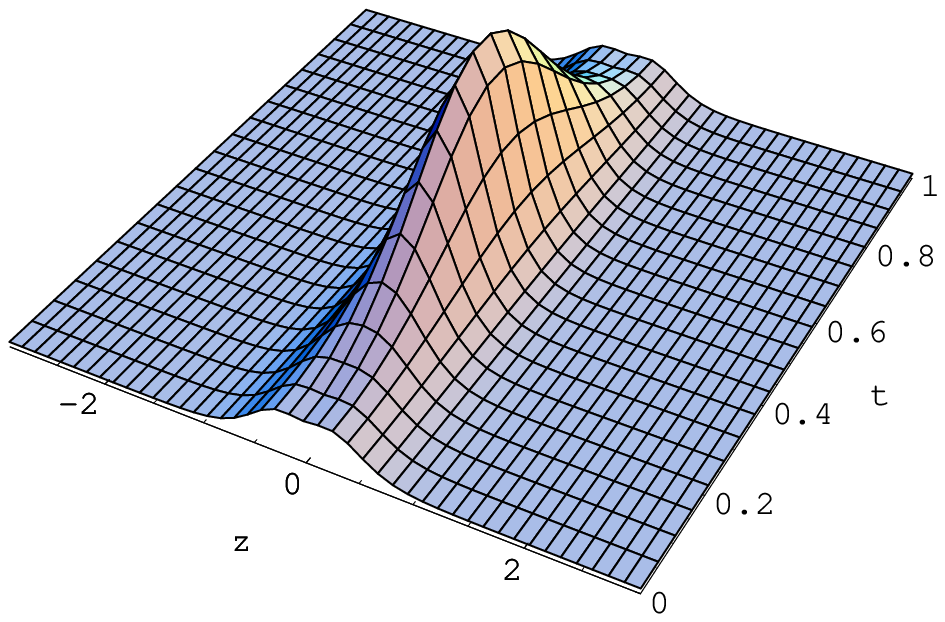}  
\caption{Action density inside the $SU(3)$ KvBLL instanton as function of time and
one space coordinate, for large ({\it left}), intermediate ({\it middle}) and small
({\it right}) separations between the three constituent dyons. }
\end{figure}

When all dyons are far apart one observes $N$ static ({\it i.e.} time-independent)
objects, the isolated dyons. As they merge, the configuration is not static anymore,
it becomes a {\it process} in time. In the limiting case of a complete merger, the
configuration becomes a $4d$ lump resembling the standard instanton. The full
(integrated) action is exactly the same $S_{\rm inst}=8\pi^2/g^2$ for any choice of
the dyon separations. It means that classically dyons do not interact. However, they
do experience a peculiar interaction at the quantum level to which we proceed.

\section{Quantum weight of a neutral cluster of $N$ dyons}

Remarkably, the small-oscillation determinant about the KvBLL instanton can be computed
exactly; this has been first done for the $SU(2)$ group in Ref.~\cite{DGPS} and later
generalized to $SU(N)$ in Ref.~\cite{S}. The quantum weight of the KvBLL instanton
can be schematically written as an integral over $3N$ coordinates of dyons (the
weight does not depend on the $U(1)$ angles $\psi_m$, hence they can be integrated out):
\beq
W_1 = \int\!d{\bf x}_1...d{\bf x}_N\,\sqrt{\det g}\,\left(\frac{4\pi}{g^4}\frac{\mu^4}{T}\right)^N\,
\exp\left(-\frac{8\pi^2}{g^2}\right)\,\left(\Det(-\triangle)\right)^{-1}_{\rm reg,\,norm},
\la{W10}\eeq
where $g$ is the full $4N\times 4N$ metric tensor of the moduli space, defined as
the zero modes overlap matrix, and $\Det(-\triangle)$ is the functional determinant over
nonzero modes, normalized to the free one and regularized by the background Pauli--Villars
method; $\mu$ is the Pauli--Villars ultra-violet cutoff and $g^2$ is the bare coupling
constant defined at that cutoff. The Jacobian $\det g$ turns out to be a square of
the determinant of an $N\times N$ matrix $G^{(1)}$ such that $\sqrt{\det g}=\det G^{(1)}$ where
\bea\la{G1}
G_{mn}^{(1)}&=&\delta_{mn}\,\left(4\pi\nu_m+\frac{1}{T|{\bf x}_m-{\bf x}_{m-1}|}
+\frac{1}{T|{\bf x}_m-{\bf x}_{m+1}|}\right)\\
\n
&&\!\!\!-\frac{\delta_{m,n-1}}{T|{\bf x}_m-{\bf x}_{m+1}|}
-\frac{\delta_{m,n+1}}{T|{\bf x}_m-{\bf x}_{m-1}|},\quad \nu_m=\mu_{m+1}-\mu_m,\;
\sum_{m=1}^N\nu_m=1,
\eea
is a matrix whose entries are Coulomb interactions between dyons that are nearest neighbours
in kind. The Coulomb interactions in the zero mode overlap matrix arise naturally from the
Coulomb asymptotics of the dyon field \ur{EB}, so it is quite simple to check that \Eq{G1}
is correct at large separations. A nontrivial fact is that \Eq{G1} is actually {\em exact}
for all separations between dyons, including the case when they strongly overlap like in
Fig.~2, right. This has been first conjectured by Lee, Weinberg and Yi~\cite{LWY}
and then proved to be indeed exact at all separations by a direct calculation by
Kraan~\cite{Kraan} and later checked in Ref.~\cite{DG05}. In the last paper it has been
also shown that in the limit of trivial holonomy ($\mu_m=0$) or vanishing temperature
the measure given by \Eqs{W10}{G1} reduces to the standard instanton measure written
in terms of the conventional ``center-size-orientation'', which is a rather nontrivial
but gratifying statement.

The functional determinant over nonzero modes $\Det^{-1}(-\triangle)$ together with
the classical action and the Pauli--Villars cutoff combine into the renormalized scale
parameter $\Lambda_{\rm PV}$, times a function of dyon separations, $\Lambda_{\rm PV}$
and T~\cite{DGPS,S}. It is a complicated function which, for the time being, we approximate
by its most essential part: a constant equal to $\exp\left(-P^{\rm pert}\right)$, where
$P^{\rm pert}$ is the perturbative gluon loop \ur{Ppert} in the background of a
constant field $A_4$ \ur{A4}. This part is necessarily present in $\Det^{-1}(-\triangle)$
as most of the $3d$ space outside the instanton's core is just a constant $A_4$ background,
and indeed the calculation~\cite{DGPS,S} exhibits this piece which is the only one
proportional to the 3-volume.

Therefore, we write the weight of the KvBLL instanton {\it i.e.} a neutral cluster
of $N$ different-kind dyons as
\beq
W_1 \approx \int\!d{\bf x}_1...d{\bf x}_N\,\det G_1\,f^N\,
\exp\left(-P^{\rm pert}(\mu_1,\ldots,\mu_N)\right)
\la{W1}\eeq
where the fugacity $f$ is
\beq
f=\frac{4\pi}{g^4}\,\frac{\Lambda^4}{T}={\cal O}(N^2).
\la{f}\eeq
The bare coupling constant $g^2$ in the pre-exponent is renormalized and starts to ``run''
only at the 2-loop level not considered here. Eventually, its argument will be
the largest scale in the vacuum, be it the temperature or the equilibrium density of dyons.

\section{Quantum weight of many dyons}

In the vacuum problem, one needs to use not one but ${\cal O}(V)$ number of KvBLL instantons
as the saddle point. Solutions with the topological charge greater than 2 will be hardly
ever known explicitly as their construction runs into the problem of resolving the nonlinear
Atiyah--Drinfeld--Hitchin--Manin--Nahm constraints. At present 2-instanton solutions
characterized by a nontrivial holonomy have been found~\cite{Nogradi} but it is insufficient.
Nevertheless, the moduli space {\it measure} of an arbitrary number $K$ of KvBLL instantons
can be constructed despite the lack of explicit solutions, at least in the approximation
which seems to be relevant for the large-volume thermodynamics, if not exactly.

When one takes a configuration of $K$ instantons each made of $N$ different-kind dyons
one encounters also same-kind dyons for which the metric \ur{G1} is inapplicable. However,
the case of identical dyons has been considered separately by Gibbons and Manton~\cite{GM2}.
The integration measure for $K$ identical dyons following from that work is, in our
notations,
\bea\n
W^{\rm ident} &=& \frac{1}{K!}\int\!d{\bf x}_1...d{\bf x}_K\,\det G^{\rm ident}_{K\times K}\,,\\
G^{\rm ident}_{ij} &=& \left\{\begin{array}{cc}4\pi\nu_m-\sum_{k\neq i}
\frac{2}{T|{\bf x}_i-{\bf x}_k|},& i=j,\\ \frac{2}{T|{\bf x}_i-{\bf x}_j|},
& i\neq j\end{array}\right.\,,
\la{Gident}\eea
where the identity factorial is inserted to avoid counting same configurations more
than once.

As in the case of different-kind dyons, this result for the metric can be easily obtained
at large separations from considering the asymptotics of the zero modes' overlap. However,
in contrast to the different-kind dyons, it is not possible to prove that this
expression is correct at all separations. Moreover, such an extension of \Eq{Gident} is
probably wrong. The metric for two same-kind dyons has been found exactly at all separations
by Atiyah and Hitchin~\cite{AH}: it is more complicated than what follows from \Eq{Gident}
at $K=2$ but differs from it by terms that are exponentially small at large separations~\cite{GM1}.
We shall neglect the difference and use the Gibbons--Manton metric at face value. The point
is, \Eq{Gident} imposes very strong repulsion between same-kind dyons (as does the
exact Atiyah--Hitchin metric), hence the range of the moduli space where the two metrics
differ is, statistically, not frequently visited by dyons. We do not have a proof that all
thermodynamic quantities will be computed correctly with this simplification: proving or
disproving it is an interesting and important problem. To remain on the safe side, one has
to admit today that the metric \ur{Gident} is applicable if the dyon ensemble is sufficiently
dilute, that is at high temperatures. Nevertheless, physical observables we compute
have a smooth limit even at $T\to 0$. Therefore, it may well prove to be a correct computation
at any temperatures, but this remains to be seen.

It is possible to combine the metric tensors for different-kind \ur{G1} and same-kind \ur{Gident}
dyons into one metric appropriate for the moduli space of $K_1$ dyons of kind 1,
$K_2$ dyons of kind 2,..., $K_N$ dyons of kind $N$. It is a matrix whose dimension is
the total number of dyons, that is a $(K_1+\ldots +K_N)\times(K_1+\ldots +K_N)$ matrix:
\bea\la{G}
G_{mi,nj}&=&\delta_{mn}\delta_{ij}\,\left(4\pi\nu_m \right.\\
\n
&+&\left.\sum_k\frac{1}{T|{\bf x}_{mi}\!-\!{\bf x}_{m-1,k}|}
+\sum_k\frac{1}{T|{\bf x}_{mi}\!-\!{\bf x}_{m+1,k}|}
-2\sum_{k\neq i}\frac{1}{T|{\bf x}_{mi}\!-\!{\bf x}_{mk}|}\right)\\
\n
&-&\frac{\delta_{m,n-1}}{T|{\bf x}_{mi}\!-\!{\bf x}_{m+1,j}|}
-\frac{\delta_{m,n+1}}{T|{\bf x}_{mi}\!-\!{\bf x}_{m-1,j}|}
+2\left.\frac{\delta_{mn}}{T|{\bf x}_{mi}\!-\!{\bf x}_{mj}|}\right|_{i\neq j},
\eea
where ${\bf x}_{mi}$ is the coordinate of the $i^{\rm th}$ dyon of kind $m$.
Since the statistical weight of a configuration of dyons is large when $\det G$
is large and small when it is small, $\det G$ imposes an attraction between
dyons that are nearest neighbours in kind, and a repulsion between same-kind dyons.
The coefficients -1,2,-1 in front of the Coulomb interactions are actually the
scalar products of the Cartan generators that determine the asymptotics of the
dyons' field, see \Eq{EB}.

The matrix $G$ has the following nice properties:
\begin{itemize}
\item symmetry: $G_{mi,nj}=G_{nj,mi}$
\item overall ``neutrality'': the sum of Coulomb interactions in non-diagonal entries
cancel those on the diagonal:
$\sum_{nj}G_{mi,nj}=4\pi\nu_m$
\item identity loss: dyons of the same kind are indistinguishable, meaning
mathematically that $\det G$ is symmetric under permutation of any pair of dyons
$(i\!\leftrightarrow\!j)$ of the same kind $m$. Dyons do not `know' to which
instanton they belong to
\item factorization: in the geometry when dyons fall into $K$ well separated
neutral clusters of $N$ dyons of different kinds in each, $\det G$ factorizes
into a product of exact integration measures for $K$ KvBLL instantons,
$\det G=(\det G^{(1)})^K$ where $G^{(1)}$ is given by \eq{G1}
\item last but not least, the metric corresponding to $G$ is hyper-K\"ahler,
as it should be for the moduli space of a self-dual classical field~\cite{AH}.
In fact, it is a severe restriction on the metric.
\end{itemize}

\section{Ensemble of dyons}

In the semiclassical approximation we thus replace the YM partition function
\ur{Z2} by the partition function of an interacting ensemble of an arbitrary number
of dyons of $N$ kinds:
\beq
{\cal Z}=\sum_{K_1...K_N}\frac{1}{K_1!...K_N!}\prod_{m=1}^N\,\prod_{i=1}^{K_m}
\int (d{\bf x}_{mi}\,f)\,\det G({\bf x}),
\la{Z3}\eeq
where ${\bf x}_{mi}$ is the coordinate of the $i^{\rm th}$ dyon of kind $m$,
the matrix $G$ is given by \Eq{G} and the fugacity $f$ is given by \Eq{f}.
The overall exponent of the perturbative potential energy as function of the
holonomy $\{\mu\}$ is understood, as in \Eq{W1}.

The ensemble defined by a determinant of a matrix whose dimension is the number of
particles, is not a usual one. More customary, the interaction is given by
the Boltzmann factor $\exp\left(-U_{\rm int}({\bf x}_1,\ldots)\right)$. Of course,
one can always present the determinant in that way using the identity
$\det G=\exp(\Tr\log G)\equiv \exp(-U_{\rm int})$ but the interactions will then
include three-, four-, five-... body forces. At the same time, it is precisely
the determinant form of the interaction that makes the statistical physics of dyons
an exactly solvable problem.

\section{Dyons' free energy: confining holonomy preferred}

The partition function \ur{Z3} can be computed directly and exactly, just by writing
the determinant of $G$ by definition as a sum of permutations of products of the matrix
entries. The result is astonishingly simple: all Coulomb interactions cancel exactly
after integration over dyons' positions,
provided the overall neutrality condition is satisfied, {\it viz.} $K_1=K_2=\ldots=K_N
=K$; otherwise the partition function is divergent. Therefore, the recipe for computing the
partition function is just to impose the neutrality condition and then to throw
out all Coulomb interactions! We have thus to take the product of $(4\pi\nu_m)$'s
from the diagonal of $G$:
$$
{\cal Z}=\sum_K\frac{(4\pi f V)^{KN}}{(K!)^N}\prod_{m=1}^N\nu_m^K\,.
$$
The quantity $4\pi f V$ is dimensionless and large for large volumes $V$. The sum can be
therefore computed from the saddle point in $K$ using the Stirling asymptotics for large
factorials, and we obtain
\beq
{\cal Z}=\exp\left(4\pi f V N\left(\nu_1\nu_2\ldots\nu_N\right)^{\frac{1}{N}}\right),\qquad
\nu_1+\nu_2+\ldots + \nu_N=1.
\la{Z0}\eeq
By definition, $F=-T\log {\cal Z}$ is the nonperturbative dyon-induced free energy as function
of the holonomy; for $N=2,3$ it is plotted in Fig.~1, bottom. Evidently, it has the minimum
at
\beq
\nu_1=\nu_2=\ldots=\nu_N=\frac{1}{N}
\la{nuconf}\eeq
corresponding to equidistant, that is confining values of $\mu$'s \ur{muconf}!
At the minimum, the free energy is
\beq
F_{\rm min}=-T\log {\cal Z}_{\rm min}=-4\pi f V T=-\frac{16\pi^2}{g^4}\,\Lambda^4\,V\,=\,
\frac{N^2}{4\pi^2}\,\frac{\Lambda^4}{\lambda^2}\,V,
\la{freeen}\eeq
and there are no Coulomb corrections to this result. In the last equation we have introduced
the $N$-independent 't Hooft coupling $\lambda\equiv\alpha_sN/2\pi$.

We note that the free energy is ${\cal O}(N^2)$ as expected on general $N$-counting
grounds and that it is temperature-independent. It corresponds to $\log {\cal Z}$ being
proportional to the 4-volume $V^{(4)}=V/T$, demonstrating the expected extensive behaviour
at low temperature.

\section{Statistical physics of dyons as a Quantum Field Theory}

Although the Coulomb interactions of dyons cancel exactly in the free energy, the
dyon ensemble defined by \Eq{Z3} is not a free gas but a highly correlated system.
To facilitate computing observables through correlation functions, we rewrite
\Eq{Z3} as an equivalent quantum field theory. As a byproduct, we shall also check
that the result for the free energy \ur{Z0} is correct.

To proceed to the quantum field theory description we use two mathematical tricks.\\

1. \underline{``Fermionization''} (Berezin~\cite{Berezin}). It is helpful to exponentiate
the Coulomb interactions rather than keeping them in $\det G$.
To that end one presents the determinant of a matrix as an integral over a finite
number of anticommuting Grassmann variables:
$$
\det (G_{AB})
=\int\!\prod_A d\psi_A^\dagger\,d\psi_A\,
\exp\left(\psi_A^\dagger\,G_{AB}\,\psi_B\right)\,.
$$

Now we have the two-body Coulomb interactions in the exponent and it is possible
to use the second trick presenting Coulomb interactions with the help of a functional
integral over an auxiliary boson field.\\

2. \underline{``Bosonization''} (Polyakov~\cite{Polyakov77}). One can write
\bea\n
\exp\left(\sum_{m,n}\frac{Q_mQ_n}{|{\bf x}_m-{\bf x}_n|}\right)
&=&\int D\phi\,\exp\left[-\int\!d{\bf x}\left(\frac{1}{16\pi}\partial_i\phi\partial_i\phi
+\rho\phi\right)\right]\\
\n
&=&\exp\left(\int \rho\frac{4\pi}{\triangle}\rho\right),\qquad
\rho = \sum Q_m\,\delta({\bf x}-{\bf x}_m).
\eea

After applying the first trick the ``charges'' $Q_m$ become Grassmann variables but
after applying the second one, it becomes easy to integrate them out since the square of
a Grassmann variable is zero. In fact one needs $2N$ boson fields $v_m,w_m$ to reproduce
diagonal elements of $G$ and $2N$ anticommuting (``ghost'') fields $\chi^\dagger_m,\chi_m$
to present the non-diagonal elements. The chain of identities is accomplished in
Ref.~\cite{DP-07} and the result for the partition function \ur{Z3} is, identically,
a path integral defining a quantum field theory in 3 dimensions:
\bea\la{Z4}
{\cal Z}&=&\int\!D\chi^\dagger\,D\chi\,Dv\,Dw\,\exp\int\!d^3x
\left\{\frac{T}{4\pi}\,\left(\partial_i\chi_m^\dagger\partial_i\chi_m
+\partial_iv_m\partial_iw_m\right)\right.\\
\n
&+&\left.f\left[(-4\pi\mu_m+v_m)\frac{\partial{\cal F}}{\partial w_m}
+\chi^\dagger_m\,\frac{\partial^2{\cal F}}{\partial w_m\partial w_n}\,\chi_n\right]
\right\}\,,\\
\la{Todapot}
{\cal F}&=&\sum_{m=1}^N e^{w_m-w_{m+1}}\,.
\eea
The fields $v_m$ have the meaning of the asymptotic Abelian electric potentials of dyons,
\bea\n
\left(A_4\right)_{mn}&=&\delta_{mn}\,A_{m\,4},\\
A_{m\,4}({\bf x})/T &=& 2\pi \mu_m-\half v_m({\bf x}),\qquad {\bf E}_m={\bf \nabla}A_{m\,4},
\la{Av}\eea
while $w_m$ have the meaning of the dual (or magnetic) Abelian potentials.
Note that the kinetic energy for the $v_m,w_m$ fields has only the mixing term
$\partial_iv_m\partial_iw_m$ which is nothing but the Abelian duality transformation
${\bf E}\cdot{\bf B}$. The function ${\cal F}(w)$ \ur{Todapot} where one assumes
a cyclic summation over $m$, is known as the periodic (or affine) Toda lattice.

Although the Lagrangian in \Eq{Z4} describes a highly nonlinear interacting quantum field
theory, it is in fact exactly solvable! To prove it, one observes that the fields $v_m$
enter the Lagrangian only linearly, therefore one can integrate them out. It leads
to a functional $\delta$-function:
\beq
\int\!Dv_m\quad\longrightarrow\quad \delta\left(-\frac{T}{4\pi}\partial^2w_m
+f\frac{\partial{\cal F}}{\partial w_m}\right).
\la{UD}\eeq
This $\delta$-function restricts possible fields $w_m$ over which one still has
to integrate in \eq{Z4}. Let $\bar w_m$ be a solution to the argument of the
$\delta$-function. Integrating over small fluctuations about $\bar w$ gives
the Jacobian
\beq\la{Jac}
{\rm Jac}={\rm det}^{-1}\left(-\frac{T}{4\pi}\partial^2\delta_{mn}
+\left.f\frac{\partial^2{\cal F}}{\partial w_m\partial w_n}\,
\right|_{w=\bar w}\,\right)\,.
\eeq
Remarkably, exactly the same functional determinant (but in the numerator)
arises from integrating over the ghost fields, for any background $\bar w$.
Therefore, all quantum corrections cancel {\em exactly} between the boson and
ghost fields (a characteristic feature of supersymmetry), and the ensemble of dyons
is basically governed by a classical field theory.

To find the ground state we examine the fields' potential energy being
$-4\pi f\mu_m\partial{\cal F}/\partial w_m$ which we prefer to write restoring
$\nu_m=\mu_{m\!+\!1}-\mu_m$ and ${\cal F}$ as
\beq
{\cal P}=-4\pi f V\sum_m \nu_m\,e^{w_m-w_{m\!+\!1}}
\la{calP1}\eeq
(the volume factor arises for constant fields $w_m$). One has first to find the
stationary point in $w_m$ for a given set of $\nu_m$'s. It leads to the equations
$$
\frac{\partial {\cal P}}{\partial w_m}=0
$$
whose solution is
\beq
e^{w_1-w_2}=\frac{(\nu_1\nu_2\nu_3...\nu_N)^{\frac{1}{N}}}{\nu_1},\quad
e^{w_2-w_3}=\frac{(\nu_1\nu_2\nu_3...\nu_N)^{\frac{1}{N}}}{\nu_2},\quad
{\rm etc.}
\la{extr2}\eeq
Putting it back into \eq{calP1} we obtain
\beq
{\cal P}=-4\pi f V N (\nu_1\nu_2...\nu_N)^{\frac{1}{N}},\qquad
\nu_1+\nu_2+...+\nu_N=1,
\la{calP2}\eeq
which is exactly what one gets from a direct calculation of the partition function,
outlined in the previous section, see \Eq{Z0}. The minimum is achieved at the
equidistant, confining value of the holonomy, see \Eqs{nuconf}{muconf}. Using field-theoretic
methods, we have also proven that the result is exact, as all potential quantum corrections
cancel. It is in line with the exact cancelation of the Coulomb interactions in the
determinant.

Given this cancelation, the key finding -- that the dyon-induced free energy has the
minimum at the confining value of holonomy -- is trivial. If all Coulomb interactions cancel
after integration over dyons' positions,
the weight of a many-dyon configuration is the same as if they were infinitely dilute
(although they are not). Then the weight, what concerns the holonomy, is proportional to
the product of diagonal matrix elements of $G$ in the dilute limit, that is to the
normalization integrals for dyon zero modes. These are nothing but the field strengths
$F_{\mu\nu}$ of individual dyons, hence the normalization is proportional to the product of
the dyon actions $\sim \nu_m$ where $\nu_m=\mu_{m+1}-\mu_m$ and $\nu_N=\mu_1+1-\mu_N$
such that $\nu_1+\nu_2+\ldots+\nu_N=1$. The sum of all $N$ kinds of dyons' actions is
fixed and equal to the instanton action, however, it is the {\em product} of actions
that defines the weight. The product is maximal when all actions are equal, hence
the equidistant or confining $\mu$'s are statistically preferred. Thus, the average
Polyakov line is zero, $<\Tr L>=0$.

\section{Heavy quark potential}

The field-theoretic representation of the dyon ensemble enables one to compute
various Yang--Mills correlation functions in the semiclassical approximation.
The key observables relevant to confinement are the correlation function
of two Polyakov lines (defining the heavy quark potential), and the average of
large Wilson loops. A detailed calculation of these quantities is performed in Ref.~\cite{DP-07};
here we only present the results and discuss the meaning.

\subsection{$N$-ality and $k$-strings}

From the viewpoint of confinement, all irreducible representations of the $SU(N)$
group fall into $N$ classes: those that appear in the direct product of any number
of adjoint representations, and those that appear in the direct product of any
number of adjoint representations with the irreducible representation being the
rank-$k$ antisymmetric tensor, $k=1,\ldots , N\!-\!1$. ``$N$-ality'' is said
to be zero in the first case and equal to $k$ in the second. $N$-ality-zero
representations transform trivially under the center of the group $Z_N$;
the rest acquire a phase $2\pi k/N$.

One expects that there is no asymptotic linear potential between static color
sources in the adjoint representation as such sources are screened by gluons.
If a representation is found in a direct product of some number of adjoint
representations and a rank-$k$ antisymmetric representation, the adjoint ones
``cancel out'' as they can be all screened by an appropriate number of gluons.
Therefore, from the confinement viewpoint all $N$-ality $=k$ representations are
equivalent and there are only $N-1$ string tensions $\sigma_{k,N}$ being the
coefficients in the {\em asymptotic} linear potential for sources in the antisymmetric
rank-$k$ representation. They are called ``$k$-strings''. The representation
dimension is $d_{k,N}=\frac{N!}{k!(N-k)!}$ and the eigenvalue of the quadratic
Casimir operator is $C_{k,N}=\frac{N+1}{2N}\,k(N-k)$.

The value $k\!=\!1$ corresponds to the fundamental representation whereas
$k=N\!-\!1$ corresponds to the representation conjugate to the fundamental
[quarks and anti-quarks]. In general, the rank-$(N\!-\!k)$ antisymmetric
representation is conjugate to the rank-$k$ one; it has the same dimension
and the same string tension, $\sigma_{k,N}=\sigma_{N\!-\!k,N}$. Therefore, for
odd $N$ all string tensions appear in equal pairs; for even $N$, apart from
pairs, there is one privileged representation with $k=\frac{N}{2}$ which
has no pair and is real. The total number of different string tensions
is thus $\left[\frac{N}{2}\right]$.

The behaviour of $\sigma_{k,N}$ as function of $k$ and $N$ is an important issue as
it discriminates between various confinement mechanisms. On general $N$-counting
grounds one can only infer that at large $N$ and $k\ll N$, $\sigma_{k,N}/\sigma_{1,N}=
(k/N)(1+{\cal O}(1/N^2))$. Important, there should be no ${\cal O}(N^{-1})$
correction~\cite{Shifman}. A popular version called ``Casimir scaling'', according
to which the string tension is proportional to the Casimir operator for a
given representation (it stems from an idea that confinement is somehow related
to the modification of a one-gluon exchange at large distances), does not satisfy
this restriction.

\subsection{Correlation function of Polyakov lines}

To find the potential energy $V_{k,N}$ of static ``quark'' and ``antiquark''
transforming according to the antisymmetric rank-$k$ representation, one has
to consider the correlation of Polyakov lines in the appropriate representation:
\beq
\left<\Tr L_{k,N}({\bf z}_1)\;\;\Tr L^\dagger_{k,N}({\bf z}_2)\right>
={\rm const.}\,\exp\left(-\frac{V_{k,N}({\bf z}_1-{\bf z}_2)}{T}\right).
\la{VkN}\eeq
Far away from dyons' cores the field is Abelian and in the field-theoretic language
of \Eq{Z4} is given by \Eq{Av}. Therefore, the Polyakov line in the fundamental
representation is
\beq
\Tr L({\bf z})=\sum_{m=1}^N Z_m,\qquad Z_m=\exp\left(2\pi i\mu_m-\frac{i}{2}v_m({\bf z})\right).
\la{Lv}\eeq
In the general antisymmetric rank-$k$ representation
\beq
\Tr L_{k,N}({\bf z})=\sum_{m_1<m_2<...<m_k}^NZ_{m_1}Z_{m_2}...Z_{m_k}
\la{LkNv}\eeq
where cyclic summation from 1 to $N$ is assumed.

The average \ur{VkN} can be computed from the quantum field theory \ur{Z4}.
Inserting the two Polyakov lines \ur{LkNv} into \Eq{Z4} we observe that
the Abelian electric potential $v_m$ enters linearly in the exponent as before.
Therefore, it can be integrated out, leading to a $\delta$-function for the
dual field $w_m$, which is now shifted by the source (cf. \Eq{UD}):
\bea\n
\int\!Dv_m &\longrightarrow & \prod_m\delta\left(-\frac{T}{4\pi}\partial^2w_m+f\frac{\partial{\cal F}}{\partial w_m}
-\frac{i}{2}\,\delta({\bf x}\!-\!{\bf z_1})(\delta_{mm_1}+\ldots +\delta_{mm_k})\right.\\
&&\left.+\frac{i}{2}\,\delta({\bf x}\!-\!{\bf z_2})(\delta_{mn_1}+\ldots +\delta_{mn_k})\right).
\la{UDL}\eea
One has to find the dual field $w_m({\bf x})$ nullifying the argument of this
$\delta$-function, plug it into the action
\beq
\exp\left(\!\int\!d{\bf x}\,\frac{4\pi f}{N}{\cal F}(w)\right),
\la{action}\eeq
and sum over all sets $\{m_1<m_2<...<m_k\}$, $\{n_1<n_2<...<n_k\}$ with the weight
$\exp\left(2\pi i(m_1+\ldots +m_k-n_1-\ldots - n_k)/N\right)$. The Jacobian from resolving
the $\delta$-function again cancels exactly with the determinant arising from ghosts.
Therefore, the calculation of the correlator \ur{VkN}, sketched above, is exact.

At large separations between the sources $|{\bf z}_1-{\bf z}_2|$, the fields $w_m$
resolving the $\delta$-function are small and one can expand the Toda chain:
\beq
{\cal F}(w)=\sum_m e^{w_m-w_{m\!+\!1}}\approx N+\frac{1}{2}\, w_m\,{\cal M}_{mn}\,w_n,
\qquad \frac{\partial {\cal F}}{\partial w_m}\approx {\cal M}_{mn}\,w_n,
\la{calFsm}\eeq
where
\beq
{\cal M}=\left(\begin{array}{cccccc}2&-1&0&\ldots&0 & -1\\ -1&2&-1&\ldots &0& 0\\
0&-1&2&-1&\ldots &0\\ \ldots & \ldots & \ldots & \ldots & \ldots & \ldots\\-1&0&0&\ldots
&-1&2\end{array}\right).
\la{calM}\eeq
As apparent from \Eq{calFsm}, the eigenvalues of ${\cal M}$ determine the spectrum of the
dual fields $w_m$. There is one zero eigenvalue which decouples from everywhere,
and $N\!-\!1$ nonzero eigenvalues
\beq
{\cal M}^{(k)}=\left(2\sin\frac{\pi k}{N}\right)^2,\quad k=1,...,N-1.
\la{eig2}\eeq
Certain orthogonality relation imposes the selection rule: the asymptotics of the correlation
function of two Polyakov lines in the antisymmetric rank-$k$ representation is determined
by precisely the  $k^{\rm th}$ eigenvalue. We obtain~\cite{DP-07}
\beq
\left<\Tr L_{k,N}({\bf z}_1)\;\;\Tr L^\dagger_{k,N}({\bf z}_2)\right>
\stackrel{z_{12}\to\infty}{=}{\rm const.}
\exp\left(-|{\bf z}_1-{\bf z}_2|\,M\sqrt{{\cal M}^{(k)}}\right)
\la{corrLk2}\eeq
where $M$ is the `dual photon' mass,
\beq
M^2=\frac{4\pi f}{T}=\frac{16\pi^2\Lambda^4}{g^4 T^2}={\cal O}(N^2).
\la{M}\eeq
Comparing it with the definition of the heavy quark potential \ur{VkN} we find
that there is an asymptotically linear potential between static ``quarks'' in
any $N$-ality nonzero representation, with the $k$-string tension
\beq
\sigma_{k,N}=MT\sqrt{{\cal M}^{(k)}}=2MT\,\sin\frac{\pi k}{N}
=\frac{\Lambda^2}{\lambda}\,\frac{N}{\pi}\,\sin\frac{\pi k}{N}.
\la{sigma-k}\eeq
This is the so-called `sine regime': it has been found before in certain supersymmetric
theories~\cite{sine}. Lattice simulations~\cite{DelDebbio-k} support this regime,
whereas another lattice study~\cite{Teper-k} gives somewhat smaller values
but within two standard deviations from the values following from \eq{sigma-k}.
For a general discussion of the sine regime for $k$-strings, which is favored
from many viewpoints, see~\cite{Shifman}.

We see that at large $N$ and $k\ll N$, $\sigma_{k,N}/\sigma_{1,N}=
(k/N)(1+{\cal O}(1/N^2))$, as it should be on general grounds, and that
all $k$-string tensions have a finite limit at zero temperature.


\section{Area law for large Wilson loops}

When dealing with the ensemble of dyons, it is convenient to use a gauge
where $A_4$ is diagonal ({\it i.e.} Abelian). This necessarily
implies Dirac string singularities sticking from dyons, which are however gauge
artifacts as they do not carry any energy. Moreover, the Dirac strings' directions
are also subject to the freedom of the gauge choice. For example, one can
choose the gauge in which $N$ dyons belonging to a neutral cluster are connected
by Dirac strings. This choice is, however, not convenient for the ensemble
as dyons have to loose their ``memory'' to what particular instanton they belong to.
The natural gauge is where all Dirac strings of all dyons are directed to infinity
along some axis, {\it e.g.} along the $z$ axis. The dyons' field in this gauge
is given explicitly in Ref.~\cite{DP-SUSY} (for the $SU(2)$ group).

In this gauge, the magnetic field of dyons beyond their cores is Abelian
and is a superposition of the Abelian fields of individual dyons. For large
Wilson loops we are interested in, it is this superposition field of a large
number of dyons that contributes most as they have a slowly decreasing
$1/|{\bf x}\!-\!{\bf x}_i|$ asymptotics, hence the use of the field outside
the cores is justified. Owing to self-duality,
\beq
\left[B_i({\bf x})\right]_{mn}=\left[\partial_iA_4({\bf x})\right]_{mn}
=-\frac{T}{2}\,\delta_{mn}\,\partial_iv_m({\bf x}),
\la{B}\eeq
cf. \eq{Av}. Since $A_i$ is Abelian beyond the cores, one can use the Stokes
theorem for the spatial Wilson loop:
\beq
W\equiv\Tr\,{\cal P}\exp\,i\oint\!A_idx^i=\Tr\exp\,i\int\!B_i\,d^2\sigma^i
=\sum_m\exp\left(-i\frac{T}{2}\int\!d^2\sigma^i\,\partial_iv_m\right).
\la{Wi1}\eeq
\Eq{Wi1} may look contradictory as we first use $B_i={\rm curl}A_i$ and then
$B_i=\partial_iA_4$. Actually there is no contradiction as the last equation
is true up to Dirac string singularities which carry away the magnetic flux.
If the Dirac string pierces the surface spanning the loop it gives a quantized
contribution $\exp(2\pi i\!\cdot\!{\rm integer})=1$; one can also use the gauge
freedom to direct Dirac strings parallel to the loop surface in which case
there is no contribution from the Dirac strings at all.

Let us take a flat Wilson loop lying in the $(xy)$ plane at $z\!=\!0$. Then
\eq{Wi1} is continued as
\bea\n
W&=&\sum_m\exp\left(-i\frac{T}{2}\int_{x,y\in {\rm Area}}\!d^3x\,\partial_zv_m\delta(z)\right)\\
&=&\sum_m\exp\left(i\frac{T}{2}\int_{x,y\in {\rm Area}}\!d^3x\,v_m\,\partial_z\delta(z)\right)\,.
\la{Wi2}\eea
It means that the average of the Wilson loop in the dyons ensemble is given by
the partition function \ur{Z4} with the source
$$
\sum_m\exp\left(i\frac{T}{2}\int\!d^3x\,v_m\,\frac{d\delta(z)}{dz}\,
\theta(x,y\in {\rm Area})\right)
$$
where $\theta(x,y\in {\rm Area})$ is a step function equal to unity if $x,y$ belong
to the area inside the loop and equal to zero otherwise. As in the case of the
Polyakov lines the presence of the Wilson loop shifts the argument of the
$\delta$-function arising from the integration over the $v_m$ variables, and the
average Wilson loop in the fundamental representation is given
by the equation
\bea\n
\left<W\right>&=&\sum_{m_1}\int\!Dw_m\exp\left(\!\int\!d{\bf x}\,\frac{4\pi f}{N}{\cal F}(w)\right)
\,\det\left(\!-\frac{T}{4\pi}\partial^2\delta_{mn}+
f\frac{\partial^2{\cal F}}{\partial w_m\partial w_n}\right)\\
\la{Wi3}
&\cdot\!&\!\!\!\!\!\prod_m
\delta\left(\!-\frac{T}{4\pi}\partial^2w_m+f\frac{\partial{\cal F}}{\partial w_m}
+\frac{i T}{2}\,\delta_{mm_1}\,\frac{d\delta(z)}{dz}\,\theta(x,y\in {\rm Area})
\right).
\eea
Therefore, one has to solve the non-linear equations on $w_m$'s with
a source along the surface of the loop,
\beq
-\partial^2w_m+M^2\left(e^{w_m-w_{m+1}}-e^{w_{m-1}-w_m}\right)
=-2\pi i\,\delta_{mm_1}\,\frac{d\delta(z)}{dz}\,\theta(x,y\in {\rm Area}),
\la{UD3}\eeq
for all $m_1$, plug it into the action $(4\pi f/N){\cal F}(w)$, and sum over
$m_ 1$. In order to evaluate the average of the Wilson loop in a general antisymmetric
rank-$k$ representation, one has to take the source in \eq{UD3} as $-2\pi i\,\delta'(z)\,
\left(\delta_{mm_1}\!+\!\ldots\!+\!\delta_{mm_k}\right)$ and sum over $m_1\!<\!\ldots\!<\!m_k$
from 1 to $N$, see \eq{LkNv}. Again, the ghost determinant cancels exactly the Jacobian from
the fluctuations of $w_m$ about the solution, therefore the classical-field calculation
is exact.

Contrary to the case of the Polyakov lines, one cannot, generally speaking,
linearize \eq{UD3} in $w_m$ but has to solve the non-linear equations as they are.
The Toda equations \ur{UD3} with a $\delta'(z)$ source in the r.h.s. define
``pinned soliton'' solutions $w_m(z)$ that are $1d$ functions in the direction
transverse to the surface spanning the Wilson loop but do not depend on the
coordinates $x,y$ provided they are taken inside the loop. Beyond that surface $w_m=0$.
Along the perimeter of the loop, $w_m$ interpolate between the soliton and zero.
For large areas, the action \ur{action} is therefore proportional to the area of the surface
spanning the loop, which gives the famous area law for the average Wilson loop.
The coefficient in the area law, the `magnetic' string tension, is found
from integrating the action on the solution $w_m(z)$ in the $z$ direction.

The exact solutions of \Eq{UD3} for any $N$ and any representation $k$ have been
found in Ref.~\cite{DP-07}, and the resulting `magnetic' string tension turns out to be
\beq
\sigma_{k,N}=\frac{\Lambda^2}{\lambda}\,\frac{N}{\pi}\,\sin\frac{\pi k}{N}\,,
\la{sigmaM-k}\eeq
which coincides with the `electric' string tension \ur{sigma-k} found
from the correlators of the Polyakov lines, for all $k$-strings!

Several comments are in order here.
\begin{itemize}
\item The `electric' and `magnetic' string tensions should coincide only in the
limit $T\to 0$ where the Euclidean $O(4)$ symmetry is restored. Both calculations
have been in fact performed in that limit as we have ignored the temperature-dependent
perturbative potential \ur{Ppert}. If it is included, the `electric' and `magnetic'
string tensions split.
\item despite that the theory \ur{Z4} is 3-dimensional, with the temperature entering
just as a parameter in the Lagrangian, it ``knows'' about the restoration of
Euclidean $O(4)$ symmetry at $T\to 0$.
\item the `electric' and `magnetic' string tensions are technically obtained in
very different ways: the first is related to the mass of the elementary excitation
of the dual fields $w_m$, whereas the latter is related to the mass of the dual field
soliton.
\end{itemize}

\section{Cancelation of gluons in the confinement phase}

To prove confinement, it is insufficient to demonstrate the area law for large
Wilson loops and the zero average for the Polyakov line: it must be shown that
there are no massless gluons left in the spectrum. We give an argument
that this indeed happens in the dyon vacuum.

A manifestation of massless gluons in perturbation theory is the Stefan--Boltzmann
law for the free energy:
\beq
-\frac{T}{V}\log {\cal Z}_{\rm SB} =\frac{F_{\rm SB}}{V}
=-\frac{\pi^2}{45}\,T^4\,(N^2-1).
\la{FSB}\eeq
It is proportional to the number of gluons $N^2\!-\!1$ and has the $T^4$ behaviour
characteristic of massless particles. In the confinement phase, neither is permissible:
If only glueballs are left in the spectrum the free energy must be ${\cal O}(N^0)$
and the temperature dependence must be very weak until $T\approx T_c$ where it abruptly
rises owing to the excitation of many glueballs.

As explained in Section 8, the ensemble of dyons has a nonperturbative free energy
\beq
\frac{F_{\rm dyon}}{V}=-\frac{N^2}{2\pi^2}\,\frac{\Lambda^4}{\lambda^2}.
\la{Fdyon}\eeq
It is ${\cal O}(N^2)$ but temperature-independent. We have doubled $F_{\rm min}$ from \Eq{freeen}
keeping in mind that there are also anti-dyons and assuming that their interactions
with dyons is not as strong as the interactions between dyons and anti-dyons
separately, as induced by the determinant measure \ur{G}, therefore treating
dyons and anti-dyons as two independent ``liquids''. (By the same logic, the string
tension \ur{sigma-k} has to be multiplied by $\sqrt{2}$ as due to anti-dyons.)

Dyons force the system to have the ``maximally nontrivial'' holonomy \ur{muconf}.
For that holonomy, the perturbative potential energy \ur{Ppert} is at its maximum equal to
\beq
\frac{F_{\rm pert,\, max}}{V}=\frac{\pi^2}{45}\,T^4\,\left(N^2-\frac{1}{N^2}\right).
\la{Fmax}\eeq
The full free energy is the sum of the three terms above.

We see that the leading ${\cal O}(N^2)$ term in the Stefan--Boltzmann law is {\em canceled}
by the potential energy precisely at the confining holonomy point and nowhere else!
In fact it seems to be the only way how ${\cal O}(N^2)$ massless gluons can be
canceled out of the free energy, and the main question shifts to why does the
system prefer the ``maximally nontrivial'' holonomy. Dyons answer that question.

\section{Deconfinement phase transition}

As the temperature rises, the perturbative free energy grows as $T^4$ and eventually it
overcomes the negative nonperturbative free energy \ur{Fdyon}. At this point, the trivial
holonomy for which both the perturbative and nonperturbative free energy are zero,
becomes favourable. Therefore an estimate of the critical deconfinement temperature
comes from equating the sum of \Eq{Fdyon} and \Eq{Fmax} to zero, which gives
\beq
T_c^4=\frac{45}{2\pi^4}\,\frac{N^4}{N^4-1}\,
\frac{\Lambda^4}{\lambda^2}\,.
\la{Tc}\eeq
As expected, it is stable in $N$. A more robust quantity, both from the theoretical
and lattice viewpoints, is the ratio $T_c/\sqrt{\sigma}$ since in this ratio
the poorly known parameters $\Lambda$ and $\lambda$ cancel out:
\beq
\frac{T_c}{\sqrt{\sigma}}=\left(\frac{45}{4\pi^4}\,\frac{\pi^2 N^2}{(N^4-1)\sin^2\frac{\pi}{N}}
\right)^{\frac{1}{4}}\quad\stackrel{N\to\infty}{\longrightarrow}\quad
\frac{1}{\pi}\left(\frac{45}{4}\right)^{\frac{1}{4}}+{\cal O}\left(\frac{1}{N^2}\right).
\la{ratio}\eeq
\vskip -0.2true cm

\begin{table}[h]
\hspace{0.4cm}
\begin{tabular}{|c|c|c|c|c|}
\hline
&$SU(3)$ & $SU(4)$ & $SU(6)$ & $SU(8)$ \\
\hline
&&&&\\
$T_c/\sqrt{\sigma}$, {\rm theory}& 0.6430 & 0.6150 & 0.5967 & 0.5906 \\
&&&&\\
\hline
&&&&\\
$T_c/\sqrt{\sigma}$, {\rm lattice}& 0.6462(30) & 0.6344(81) & 0.6101(51) & 0.5928(107) \\
&&&&\\
\hline
\end{tabular}
\end{table}

In the Table, we compare the values from \Eq{ratio} to those measured in lattice simulations
of the pure $SU(N)$ gauge theories~\cite{Teper}; there is a surprisingly good agreement.
A detailed study of the thermodynamics of the phase transition will be published elsewhere.

\section{Relation to other suggestions to explain confinement}

Several mechanisms of confinement have been suggested in the past. The most popular are
\begin{itemize}
\item condensation of monopoles, or the dual Meissner effect~\cite{tHooft-conf,Mandelstam}
\item proliferation of center vortices~\cite{tHooft-conf,Mack}, see a modern overview~\cite{Greensite}
\end{itemize}
These two mechanisms and in particular lattice evidence supporting them have been reviewed
by Jeff Greensite~\cite{Greensite} and we are not going to repeat it here. What is important,
both monopoles and vortices are identified on a lattice by fixing the gauge -- choosing
the ``maximally Abelian'' gauge in the first case and the ``maximally center'' gauge
in the second. If this gauge-fixing procedure is applied to the dyon vacuum of the present
paper, the maximally Abelian gauge would probably reveal lattice monopoles where
dyons are placed, and a subsequent application of the maximally center gauge would
probably reveal center vortices which would be nothing but the phantom Dirac strings
connecting dyons. Therefore, lattice findings that ``there is no confinement without
Abelian monopoles'' and that ``there is no confinement without center vortices''
is presumably in no contradiction with the vacuum being formed by dyons. Moreover,
recently there have been direct observations of dyons on the lattice by the Humboldt
Universit\"at -- ITEP group, see~\cite{HU-ITEP} and further references therein.

Some time ago we have observed that standard instantons are also capable
of yielding confinement, provided the instanton size distribution falls off as $1/\rho^3$
at large sizes $\rho$~\cite{DP-rhocube}. This regime implies, however, that
large-size instantons inevitably overlap, since in $4d$ the packing fraction is
proportional to the fourth moment of the size distribution $\overline{\rho^4}$
which is divergent. Therefore, the usual instantons' ``center-size-orientation''
parameterization being all right for dilute systems is inapplicable for the confinement
purposes. One needs a parametrization of the collective instantons' coordinates
that is as good for overlapping solutions as it is for dilute ones.

In an analogous $2d$ $CP^{N\!-\!1}$ model also possessing instantons such a parameterization
has long been known: instantons there are parameterized by the positions of $N$ kinds of
``instanton quarks''. The measure of the moduli space of multi-instantons is fortunately
known exactly~\cite{FFS-BL} and is given by a holomorphic function of the instanton quarks'
coordinates. The measure is invariant under permutation of the instanton quarks
(they should not `know' what instanton they belong to) and is perfectly valid for overlapping
instantons, as well as for dilute ones. In the latter case the measure becomes
the product of instanton ``center-size-orientation'' measures~\cite{DM}.

In the $4d$ YM theory a similar parameterization of multi-instantons has long been
sought, starting from the pioneering work of Callan, Dashen and Gross who suggested
``merons'' as instanton constituents~\cite{CDG}, but that did not work as merons had
a divergent action. Zhitnitsky~\cite{Zhitnitsky}, Petrov and myself put much effort
in identifying ``instanton quarks'' for the YM solutions but real progress has been
achieved in constructing the KvBLL instantons~\cite{KvB,LL} whose
constituents have been found to be the BPS monopoles, or dyons. The price is that
one is obliged to take nonzero temperatures, however if one is interested in the
zero-temperature case, $T$ can be considered as an infrared regulator which is safe
to put to zero at the end, if needed.

The measure of the multi-instanton space \ur{G} is now written in terms of the
coordinates of the constituent dyons. The metric is hyper-K\"ahler (which is
the $4d$ analogue of holomorphy in $2d$), the measure is invariant under permutation
of dyons (they should not `know' what instanton they belong to) and is presumably valid
for overlapping instantons, as well as for dilute ones. In the latter case the measure
becomes the product of the instanton ``center-size-orientation'' measures~\cite{DG05}.
Therefore, it seems to be the solution of a long-standing problem.

Two steps in modernizing the semi-classical ``instanton liquid'' model~\cite{D-02}
are critical in getting confinement:
\begin{itemize}
\item generalizing instantons in such a way that they can have arbitrary holonomy,
and allowing nontrivial holonomy, despite that in perturbation theory it is forbidden
\item writing the quantum weight of instantons with nontrivial holonomy through coordinates
of constituent dyons, such that it is applicable for overlapping instantons.
\end{itemize}

\noindent
What happens, can be summarized as follows:
\begin{itemize}
\item The ensemble of dyons favours dynamically the ``maximally nontrivial" or confining
value of the holonomy. This is almost clear, given that the weight is proportional
to the product of individual actions of $N$ kinds of dyons
\item Dyons form a sort of Coulomb plasma (but an exactly solvable variant of it)
with an appearance of the Debye mass both for ``electric'' and ``magnetic'' (dual) photons.
The first give rise to the exponential fall-off of the correlation of two Polyakov lines,
{\it i.e.} to the linear heavy-quark potential, the second yield the area law for spatial
Wilson loops
\item $N^2\!-\!1$ massless gluons cancel out from the free energy, and only massive (string?)
excitations are left.
\end{itemize}

\noindent
We do not see the quantum-mechanical condensation of monopoles; it is hence a new mechanism
of confinement.

\section{Why does it work and what should be done next?}

The reason why a semiclassical approximation works well for strong interactions
(where all dimensionless quantities are, generally speaking, of the order of unity) is
not altogether clear. A possible justification has been outlined in Section 1: After
UV renormalization is performed about the classical saddle points and the scale parameter
$\Lambda$ appears as the result of the dimensional transmutation, further quantum corrections
to the saddle point is a series in the running 't Hooft coupling $\lambda$ whose argument
is typically the largest scale in the theory, in this case ${\rm max}(T, n^{1/4})$ where
$n$ is the $4d$ density of the dyons. An estimate shows that the running $\lambda$ is
between $1/4$ at zero temperature and $1/7$ or less at critical temperature. Therefore,
although these numbers are ``of the order of unity'', in practical terms they indicate
that high order loop corrections are not too large. Let us recall that quite an accurate
computation of anomalous dimensions in critical phenomena from the $\epsilon$-expansion
by Fisher and Wilson~\cite{FW} is based on truncating the Taylor expansion in $\epsilon$
at the first couple of terms, where $\epsilon\!=\!1$ or sometimes 2 !\footnote{I take
the opportunity to thank Michael Fisher and Valery Pokrovsky for a discussion of
this numerical miracle.}

Unfortunately, approximations made in Ref.~\cite{DP-07} and reproduced above are not
limited to higher loop corrections. We have (i) ignored dyon interactions induced
by the small oscillation determinant over nonzero modes, except the potential energy
as function of the holonomy, (ii) ignored the interactions of dyons of different
duality, treating them as two noninteracting ``liquids'', (iii) conjectured a simple
form of the dyon measure which may be incorrect when two {\em same-kind} dyons
come close. Although certain justification for these approximations can be put
forward (see above and Ref.~\cite{DP-07}) it is desirable not to use them at all,
and that may be possible.

These mathematical problems are of course in the line, as well as further physical
problems, probably the most urgent being switching in light dynamical quarks into
the dyon vacuum, that is moving into the realm of the real-world QCD. The main problem there
is the spontaneous breaking of chiral symmetry. Although we do not think that its mechanism
will differ dramatically from that found in Ref.~\cite{DP-8486},
as due to the delocalization of the near-zero fermion modes, it would be very interesting
to see how the ensuing effective chiral lagrangian ``knows'' about the confinement. \\

{\bf Acknowledgements}\\

I would like to thank Nick Dorey for helpful conversations during the School in Zakopane,
and its organizers, especially Michal Prasza{\l}owicz, for most kind hospitality.
{\it Dzi\c{e}kuj\c{e} bardzo!} Almost continuous discussions with Victor Petrov,
the co-author of Ref.~\cite{DP-07} on which these notes are based, are gratefully acknowledged.
This work has been supported in part by Russian Government grants RFBR-06-02-16786
and RSGSS-5788.2006.2.

\end{document}